\documentclass[italian,english]{article}
\usepackage[T1]{fontenc}
\usepackage[latin1]{inputenc}
\usepackage{color}
\usepackage{amssymb}

\makeatletter

 \newcommand{\lyxaddress}[1]{
   \par {\raggedright #1 
   \vspace{1.4em}
   \noindent\par}
 }

\usepackage{babel}
\makeatother
\begin{document}

\title{\textbf{Moriond 07 proceedings: {}``Extension of the frequency-range
of interferometers for the ''magnetic'' components of gravitational
waves?''}}

\author{\textbf{Christian Corda}}

\maketitle

\lyxaddress{\begin{center}INFN - Sezione di Pisa and Università di Pisa, Via
F. Buonarroti 2, I - 56127 PISA, Italy\end{center}}

\lyxaddress{\begin{center}\textit{E-mail address:} \textcolor{blue}{christian.corda@ego-gw.it} \end{center}}

\begin{abstract}
Recently some papers in the literature have shown the presence and
importance of the so-called {}``magnetic'' components of gravitational
waves (GWs), which have to be taken into account in the context of
the total response functions of interferometers for GWs propagating
from arbitrary directions. In this paper the response functions for
the magnetic components are re-analysed in the full frequency dependence
answering to the question of the title.
\end{abstract}

\lyxaddress{PACS numbers: 04.80.Nn, 04.80.-y, 04.25.Nx}

The design and construction of a number of sensitive detectors for
GWs is underway today. There are some laser interferometers like the
VIRGO detector, being built in Cascina, near Pisa by a joint Italian-French
collaboration \cite{key-1,key-2}, the GEO 600 detector, being built
in Hanover, Germany by a joint Anglo-Germany collaboration \cite{key-3,key-4},
the two LIGO detectors, being built in the United States (one in Hanford,
Washington and the other in Livingston, Louisiana) by a joint Caltech-Mit
collaboration \cite{key-5,key-6}, and the TAMA 300 detector, being
built near Tokyo, Japan \cite{key-7,key-8}. There are many bar detectors
currently in operation too, and several interferometers and bars are
in a phase of planning and proposal stages.

The results of these detectors will have a fundamental impact on astrophysics
and gravitation physics. There will be lots of experimental data to
be analyzed, and theorists will be forced to interact with lots of
experiments and data analysts to extract the physics from the data
stream.

Detectors for GWs will also be important to confirm or ruling out
the physical consistency of General Relativity or of any other theory
of gravitation \cite{key-9,key-10,key-11,key-12}. This is because,
in the context of Extended Theories of Gravity, some differences from
General Relativity and the others theories can be seen starting by
the linearized theory of gravity \cite{key-9,key-10,key-12}. 

Recently some papers in the literature have shown the presence and
importance of the so-called {}``magnetic'' components of gravitational
waves (GWs), which have to be taken into account in the context of
the total response functions of interferometers for GWs propagating
from arbitrary directions \cite{key-13,key-14,key-15,key-16}. In
this paper the response functions for the magnetic components are
re-analysed in the full frequency dependence.

In a laboratory enviroment on earth, the coordinate system in which
the space-time is locally flat is typically used \cite{key-12,key-13,key-14,key-15,key-16,key-17,key-18}
and the distance between any two points is given simply by the difference
in their coordinates in the sense of Newtonian physics. In this frame,
called the frame of the local observer, GWs manifest themself by exerting
tidal forces on the masses (the mirror and the beam-splitter in the
case of an interferometer). 

The importance of the {}``magnetic'' components of GWs arises from
Section 3 of \cite{key-13}. Working with $G=1$, $c=1$ and $\hbar=1$
and calling $h_{+}(t_{tt}+z_{tt})$ and $h_{\times}(t_{tt}+z_{tt})$
the weak perturbations due to the $+$ and the $\times$ polarizations
of the GW, which are expressed in terms of syncrony coordinates $t_{tt},x_{tt},y_{tt},z_{tt}$
in the transverse-traceless (TT) gauge, the most general GW propagating
in the $z_{tt}$ direction can be written in terms of plane monochromatic
waves \cite{key-14,key-15,key-16,key-17,key-19}

\begin{equation}
\begin{array}{c}
h_{\mu\nu}(t_{tt}+z_{tt})=h_{+}(t_{tt}+z_{tt})e_{\mu\nu}^{(+)}+h_{\times}(t_{tt}+z_{tt})e_{\mu\nu}^{(\times)}=\\
\\=h_{+0}\exp i\omega(t_{tt}+z_{tt})e_{\mu\nu}^{(+)}+h_{\times0}\exp i\omega(t_{tt}+z_{tt})e_{\mu\nu}^{(\times)},\end{array}\label{eq: onda generale}\end{equation}

and the correspondent line element will be

\begin{equation}
ds^{2}=dt_{tt}^{2}-dz_{tt}^{2}-(1+h_{+})dx_{tt}^{2}-(1-h_{+})dy_{tt}^{2}-2h_{\times}dx_{tt}dx_{tt}.\label{eq: metrica TT totale}\end{equation}

The coordinate transformation $x^{\alpha}=x^{\alpha}(x_{tt}^{\beta})$
from the TT coordinates to the frame of the local observer is \cite{key-13,key-14,key-15}

\begin{equation}
\begin{array}{c}
t=t_{tt}+\frac{1}{4}(x_{tt}^{2}-y_{tt}^{2})\dot{h}_{+}-\frac{1}{2}x_{tt}y_{tt}\dot{h}_{\times}\\
\\x=x_{tt}+\frac{1}{2}x_{tt}h_{+}-\frac{1}{2}y_{tt}h_{\times}+\frac{1}{2}x_{tt}z_{tt}\dot{h}_{+}-\frac{1}{2}y_{tt}z_{tt}\dot{h}_{\times}\\
\\y=y_{tt}+\frac{1}{2}y_{tt}h_{+}-\frac{1}{2}x_{tt}h_{\times}+\frac{1}{2}y_{tt}z_{tt}\dot{h}_{+}-\frac{1}{2}x_{tt}z_{tt}\dot{h}_{\times}\\
\\z=z_{tt}-\frac{1}{4}(x_{tt}^{2}-y_{tt}^{2})\dot{h}_{+}+\frac{1}{2}x_{tt}y_{tt}\dot{h}_{\times}.\end{array}\label{eq: trasf. coord.}\end{equation}

In eqs. (\ref{eq: trasf. coord.}) it is $\dot{h}_{+}\equiv\frac{\partial h_{+}}{\partial t}$
and $\dot{h}_{\times}\equiv\frac{\partial h_{\times}}{\partial t}$.
We emphasize that, in refs. \cite{key-13,key-14,key-15,key-20} it
has been shown that the linear and quadratics terms, as powers of
$x_{tt}^{\alpha}$, are unambiguously determined by the conditions
of the frame of the local observer.

Considering a free mass riding on a timelike geodesic ($x=l_{1}$,
$y=l_{2},$ $z=l_{3}$) \cite{key-13} eqs. (\ref{eq: trasf. coord.})
define the motion of this mass with respect the introduced frame of
the local observer. If one neglects the terms with $h_{+}$and $h_{\times}$
in eqs. (\ref{eq: trasf. coord.}) the analogue of the magnetic component
of motion in electrodynamics are directly obtained:\begin{equation}
\begin{array}{c}
x(t+z)=l_{1}+\frac{1}{2}l_{1}l_{3}\dot{h}_{+}(t+z)+\frac{1}{2}l_{2}l_{3}\dot{h}_{\times}(t+z)\\
\\y(t+z)=l_{2}-\frac{1}{2}l_{2}l_{3}\dot{h}_{+}(t+z)+\frac{1}{2}l_{1}l_{3}\dot{h}_{\times}(t+z)\\
\\z(t+z)=l_{3}-\frac{1}{4[}(l_{1}^{2}-l_{2}^{2})\dot{h}_{+}(t+z)+2l_{1}l_{2}\dot{h}_{\times}(t+z),\end{array}\label{eq: news}\end{equation}

To compute the total response functions of interferometers for the
magnetic components generalized in their full frequency dependence
an analysis parallel to the one used for the first time in \cite{key-16}
has been used in \cite{key-15} the so called {}``bounching photon
metod''. We emphasize that this metod has been generalized to scalar
waves, angular dependence and massive modes of GWs in \cite{key-12}.
In \cite{key-15} it is shown that, in the frame of the local observer,
we have to consider two different effects in the calculation of the
variation of the round-trip time for photons, in analogy with the
cases of \cite{key-15} where the effects considered were three, but
the third effect vanishes putting the origin of our coordinate system
in the beam splitter of our interferometer (see also the massive case
in \cite{key-12}).

Details of computations are in \cite{key-15} , here only the response
function of the magnetic component of the {}``$+$'' polarization
will be written. 

The total response function for the magnetic component of the $+$
polarization is given by \begin{equation}
\begin{array}{c}
H_{tot}^{+}(\omega)=\frac{\tilde{\delta}T_{tot}(\omega)}{L\tilde{h}_{+}(\omega)}=\\
\\=-i\omega\exp[i\omega L(1-\sin\theta\cos\phi)]\frac{LA}{2}+\frac{LB}{2}i\omega\exp[i\omega L(1-\sin\theta\sin\phi)]\\
\\-\frac{i\omega LA}{4}[\frac{-1+\exp[i\omega L(1-\sin\theta\cos\phi)]-iL\omega(1-\sin\theta\cos\phi)}{(1-\sin\theta\cos\phi)^{2}}\\
\\+\frac{\exp(2i\omega L)(1-\exp[i\omega L(-1-\sin\theta\cos\phi)]-iL\omega(1+\sin\theta\cos\phi)}{(-1-\sin\theta\cos\phi)^{2}}]+\\
\\+\frac{i\omega LB}{4}[\frac{-1+\exp[i\omega L(1-\sin\theta\sin\phi)]-iL\omega(1-\sin\theta\sin\phi)}{(1-\sin\theta\cos\phi)^{2}}+\\
\\+\frac{\exp(2i\omega L)(1-\exp[i\omega L(-1-\sin\theta\sin\phi)]-iL\omega(1+\sin\theta\sin\phi)}{(-1-\sin\theta\sin\phi)^{2}}],\end{array}\label{eq: risposta totale 2}\end{equation}

that, in the low freuencies limit is in perfect agreement with the
result of Baskaran and Grishchuk (eq. 49 of \cite{key-13}): \begin{equation}
H_{tot}^{+}(\omega\rightarrow0)=\frac{1}{4}\sin\theta[(\cos^{2}\theta+\sin2\phi\frac{1+\cos^{2}\theta}{2})](\cos\phi-\sin\phi).\label{eq: risposta totale bassa}\end{equation}

Because the response functions to the {}``magnetic'' components
grow with frequency, as it is shown in eq. (\ref{eq: risposta totale 2})
(and the same happens for the {}``$\times$'' polarization, see
\cite{key-15}) one could think that the part of signal which arises
from the magnetic components could in principle become the dominant
part of the signal at high frequencies (see the correspondent pictures
in \cite{key-15}), but, to undesrtand if this is correct, one has
to use the full theory of gravitational waves.

The low-frequencies approximation, used in \cite{key-13} to show
that the {}``magnetic'' and {}``electric'' contributions to the
response functions can be identified without ambiguity in the longh-wavelengths
regime (see also \cite{key-13}), is sufficient only for ground based
interferometers, for which the condition $f\ll1/L$ is in general
satisfied. For space-based interferometers, for which the above condition
is not satisfied in the high-frequency portion of the sensitivity
band \cite{key-13,key-14,key-22}, the full theory of gravitational
waves has to be used.

If one removes the low-frequencies approximation, the {}``bouncing
photon method'' can be used in this context too. In \cite{key-14}
the variation of the proper distance that a photon covers to make
 a round-trip from the beam-splitter to the mirror of an interferometer
\cite{key-12,key-14,key-15,key-16} was computed with the gauge choice
(\ref{eq: metrica TT totale}). In this case one does not need the
coordinate transformation (\ref{eq: trasf. coord.}) from the TT coordinates
to the frame of the local observer (see also Section 5 of \cite{key-13}).
Even in this case, only the result for the total response function
of the {}``$+$'' polarization which takes into account both the
magnetic and electric components will be written. Details of computations
are in \cite{key-14} for the {}``$\times$'' polarization too.

Thus, the total frequency-dependent response function (i.e. the detector
pattern) of an interferometer to the $+$ polarization of the GW is:

\begin{equation}
\begin{array}{c}
\tilde{H}^{+}(\omega)=\\
\\=\frac{(\cos^{2}\theta\cos^{2}\phi-\sin^{2}\phi)}{2L}\tilde{H}_{u}(\omega,\theta,\phi)+\\
\\-\frac{(\cos^{2}\theta\sin^{2}\phi-\cos^{2}\phi)}{2L}\tilde{H}_{v}(\omega,\theta,\phi)\end{array}\label{eq: risposta totale Virgo +}\end{equation}

that, in the low frequencies limit ($\omega\rightarrow0$), if one
retains the first two terms of the expansion, is in perfect agreement
with the detector pattern of eq. (46) in \cite{key-13},:

\begin{equation}
\begin{array}{c}
\tilde{H}^{+}(\omega\rightarrow0)=\frac{1}{2}(1+\cos^{2}\theta)\cos2\phi+\\
\\-\frac{1}{4}i\omega L\sin\theta[(\cos^{2}\theta+\sin2\phi\frac{1+\cos^{2}\theta}{2})](\cos\phi-\sin\phi).\end{array}\label{eq: risposta totale approssimata}\end{equation}

This result also confirms that the magnetic contribution to the variation
of the distance is an universal phenomenon because it has been obtained
starting from the full theory of gravitational waves in the TT gauge
(see \cite{key-13,key-14}). 

Seeing the pictures of eqs. (\ref{eq: risposta totale Virgo +}) and
of the total response function of the {}``$\times$'' polarization,
which are in \cite{key-14}, one sees that the magnetic component
of GWs \textbf{cannot} extend the frequency range of interferometers.
This is because, even if magnetic contributions grow with frequency,
as it is shown from eq. (\ref{eq: risposta totale 2}), the division
between {}``electric'' and {}``magnetic'' contributions breaks
down at high frequencies, thus one has to perform computations using
the full theory of gravitational waves. The correspondent response
functions which are obtained do not grow which frequency.


\begin{thebibliography}{10}
\bibitem{key-1}Acernese F et al. (the Virgo Collaboration) - Class. Quant. Grav.
\textbf{23} 8 S63-S69 (2006) 
\bibitem{key-2}Corda C - Astropart. Phys. doi: 10.1016/ j.astropartphys.2007.04.001
(2007)
\bibitem{key-3}Hild S (for the LIGO Scientific Collaboration) - Class. Quant. Grav.
\textbf{23} 19 S643-S651 (2006)
\bibitem{key-4}Willke B et al. - Class. Quant. Grav. \textbf{23} 8S207-S214 (2006) 
\bibitem{key-5}Sigg D (for the LIGO Scientific Collaboration) - www.ligo.org/pdf\_public/P050036.pdf
\bibitem{key-6}Abbott B et al. (the LIGO Scientific Collaboration) - Phys. Rev. D
72, 042002 (2005) 
\bibitem{key-7}Ando M and the TAMA Collaboration - Class. Quant. Grav. \textbf{19}
7 1615-1621 (2002)
\bibitem{key-8}Tatsumi D, Tsunesada Y and the TAMA Collaboration - Class. Quant.
Grav. \textbf{21} 5 S451-S456 (2004) 
\bibitem{key-9}Capozziello S - \textit{Newtonian Limit of Extended Theories of Gravity}
in \textit{Quantum Gravity Research Trends} Ed. A. Reimer, pp. 227-276
Nova Science Publishers Inc., NY (2005) - \foreignlanguage{italian}{also
in arXiv:}gr-qc/0412088 (2004) 
\bibitem{key-10}Capozziello S and Troisi A - Phys. Rev. D \textbf{72} 044022 (2005) 
\bibitem{key-11}Will C M \textit{Theory and Experiments in Gravitational Physics},
Cambridge Univ. Press Cambridge (1993)
\bibitem{key-12}Capozziello S and Corda C - Int. J. Mod. Phys. D \textbf{15} 1119
-1150 (2006); Corda C - \textit{Response of laser interferometers
to scalar gravitational waves}- talk in the \textit{Gravitational
Waves Data Analysis Workshop in the General Relativity Trimester of
the Institut Henri Poincare -} Paris 13-17 November 2006, on the web
in www.luth2.obspm.fr/IHP06/workshops/gwdata/corda.pdf
\bibitem{key-13}Baskaran D and Grishchuk LP - Class. Quant. Grav. \textbf{21} 4041-4061
(2004)
\bibitem{key-14}Corda C - gr-qc/0702080 (2007)
\bibitem{key-15}Corda C - gr-qc/0610156 (2007)
\bibitem{key-16}Rakhmanov M - Phys. Rev. D \textbf{71} 084003 (2005)
\selectlanguage{italian}
\bibitem{key-17}Misner CW, Thorne KS and Wheeler JA - {}``Gravitation'' - W.H.Feeman
and Company - 1973
\selectlanguage{english}
\bibitem{key-18}Landau L and Lifsits E - {}``Teoria dei campi'' - Editori riuniti
edition III (1999)
\bibitem{key-19}Maggiore M - Physics Reports \textbf{331}, 283-367 (2000)
\bibitem{key-20}Grishchuk LP - Sov. Phys. Usp. \textbf{20} 319 (1977) 
\bibitem{key-21}Saulson P - \textit{Fundamental of Interferometric Gravitational Waves
Detectors} - World Scientific, Singapore (1994) Estabrook FB and Wahlquist
HD - Gen. Relativ. Gravit. \textbf{6} 439 (1975) Thorne KS - \textit{300
Years of Gravitation} - Ed. Hawking SW and Israel W Cambridge University
Press p. 330 (1987)
\bibitem{key-22}Estabrook FB - Gen. Relativ. Gravit. \textbf{17} 719 (1975)
\end{thebibliography}
\end{document}